\documentclass[a4paper, fleqn, usenatbib, useAMS]{mnras}
\usepackage[usenames, dvipsnames]{color}
\usepackage[T1]{fontenc}
\usepackage{ae,aecompl}
\usepackage{graphicx}
\usepackage{lscape}
\usepackage{amsmath}
\usepackage{amssymb}
\usepackage{xspace}
\usepackage{ulem}

\newcommand{\Ha}{\ifmmode {\rm H}\alpha \else H$\alpha$\fi\xspace}
\newcommand{\Hb}{\ifmmode {\rm H}\beta \else H$\beta$\fi\xspace}
\newcommand{\Hba}{\ifmmode {\rm H}\beta^{\prime} \else H$\beta^{\prime}$\fi\xspace}
\newcommand{\Hg}{\ifmmode {\rm H}\gamma \else H$\gamma$\fi\xspace}
\newcommand{\Hd}{\ifmmode {\rm H}\delta \else H$\delta$\fi\xspace}
\newcommand{\hi}{H\,{\sc i}\xspace}
\newcommand{\hii}{H\,{\sc ii}\xspace}

\newcommand{\Hii}{\ifmmode \rm{H}\,\textsc{ii} \else H\,{\sc ii}\fi}
\newcommand{\Nii}{[N\,{\sc ii}]$\lambda$6584}
\newcommand{\nii}{\ifmmode [\rm{N}\,\textsc{ii}] \else [N\,{\sc ii}]\fi\xspace}

\newcommand{\oi}{\ifmmode [\rm{O}\,\textsc{i}] \else [O\,{\sc i}]\fi\xspace}

\newcommand{\neiii}{\ifmmode [\rm{Ne}\,\textsc{iii}] \else [Ne\,{\sc iii}]\fi}

\newcommand{\hei}{\ifmmode [\rm{He}\,\textsc{i}] \else [He\,{\sc i}]\fi}
\newcommand{\oii}{\ifmmode [\rm{O}\,\textsc{ii}] \else [O\,{\sc ii}]\fi\xspace}
\newcommand{\Oiii}{[O\,{\sc iii}]$\lambda$5007}

\newcommand{\oiii}{\ifmmode [\rm{O}\,\textsc{iii}] \else [O\,{\sc iii}]\fi\xspace}
\newcommand{\Sii}{[S\,{\sc ii}]$\lambda\lambda$6731,6716}
\newcommand{\sii}{\ifmmode [\rm{S}\,\textsc{ii}] \else [S\,{\sc ii}]\fi\xspace}
\newcommand{\siii}{\ifmmode [\rm{S}\,\textsc{iii}] \else [S\,{\sc iii}]\fi}

\newcommand\pycasso{{\sc p}y{\sc casso}}          	% PyCASSO
\newcommand\starlight{{\sc starlight}}          		% STARLIGHT

\usepackage[usenames]{color}                                   % Comment this if compiled with [referee] option
\title[DIG across the Hubble sequence]{Diffuse ionized gas in galaxies across the Hubble sequence at the CALIFA resolution}

\author[Lacerda et.\ al.] {E. A. D. Lacerda$^1$\thanks{E-mail: lacerda@astro.ufsc.br},
 R. Cid Fernandes$^1$,
 G. S. Couto$^1$,
 G. Stasi\'nska$^2$,
 R. Garc\'{\i}a-Benito$^3$, \and
 N. Vale Asari$^1$,
 E. P\'erez$^3$,
 R. M. Gonz\'alez Delgado$^3$,
S. F. S\'anchez$^4$,
A. L. de Amorim$^1$ \\
 $^1$Departamento de F\'{\i}sica - CFM - Universidade Federal de Santa Catarina, P.O. Box 476, 88040-900.
 Florian\'opolis, SC, Brazil\\
 $^2$LUTH, Observatoire de Paris, CNRS, Universit\'e Paris Diderot; Place Jules Janssen 92190 Meudon, France\\
 $^3$Instituto de Astrof\'{\i}sica de Andaluc\'{\i}a (CSIC), P.O. Box 3004, 18080 Granada, Spain\\
 $^4$Instituto de Astronom\'{\i}a, Universidad Nacional Aut\'onoma de M\'exico, A. P. 70-264, C.P. 04510, M\'exico, D.F., M\'exico
}

\date{20 November 2017}

\pubyear{2017}

\begin{document}
\label{firstpage}
\pagerange{\pageref{firstpage}--\pageref{lastpage}}
\maketitle

%\pagerange{\pageref{firstpage}--\pageref{lastpage}} \pubyear{2009}

\begin{abstract}
We use spatially resolved spectroscopy from the Calar Alto Legacy Integral Field Area (CALIFA) survey to study the nature of the line emitting gas in galaxies of different Hubble types, focusing on the separation of star-forming (SF) regions from those better characterized as diffuse ionized gas (DIG). The diagnosis is carried out in terms of the equivalent width of \Ha ($W_{\Ha}$). Three nebular regimes are identified. Regions where $W_{\Ha} < 3$ \AA\  define what we call the hDIG, the component of the DIG where photoionization is dominated by hot, low-mass, evolved stars. Regions where $W_{\Ha} > 14$ \AA\ trace SF complexes. $W_{\Ha}$ values in the intermediate 3--14 \AA\ range reflect a mixed regime (mDIG) where more than one process contributes.This three-tier scheme is inspired both by theoretical and empirical considerations. Its application to CALIFA galaxies  of different types and inclinations leads to the following results: \textit{(i)} the hDIG component is prevalent throughout ellipticals and S0's as well as in bulges, and explains the strongly bimodal distribution of $W_{\Ha}$ both among and within galaxies. \textit{(ii)} Early-type spirals have some hDIG in their discs, but this component becomes progressively less relevant for  later Hubble types. \textit{(iii)}  hDIG emission is also present above and below galactic discs, as  seen in several edge-on spirals in our sample. \textit{(iv)} The SF/mDIG proportion grows steadily from early- to late-type spirals, and from inner to outer radii. \textit{(v)} Besides circumventing basic inconsistencies in conventional DIG/SF separation criteria based on the \Ha\ surface brightness, our $W_{\Ha}$-based method produces results in agreement with a classical excitation diagram analysis.

\end{abstract}


\label{firstpage}

\section{Introduction}
\label{sec:Intro}

Much of what we know about galaxies comes from their optical emission-lines. Measurements of lines like \Ha, \Hb, \oiii$\lambda5007$, \nii$\lambda6584$ and others allow properties like star formation rates, dust attenuation, gas-phase metallicity, and the presence and intensity of an active nucleus to be estimated through well-known recipes. This approach of converting emission-line data on to astrophysically valuable information has been extensively explored in the context of large surveys like the Sloan Digital Sky Survey \citep{York.etal.2000a}, producing important results in the quest for a better understanding of galaxy physics. Examples include the relation between the star formation rate and galaxy stellar mass \citep{Brinchmann.etal.2004a}, the mass--metallicity relation \citep{Tremonti.etal.2004a}, connections between AGN power and stellar populations \citep{Kauffmann.etal.2003a}, and between gas-phase metallicities and star formation histories \citep{CidFernandes.etal.2007}, to name but a few.

A fundamental caveat of integrated-light (single-fiber) spectroscopy of galaxies is that it effectively treats as a single point source what is in reality a complex system, with  physically and structurally different components. Disc and bulge, arm and inter-arm, \hii regions and diffuse gas,  dusty and clean regions, young and old stars are all mixed in a total spectrum where the parts are no longer recognizable. This simplified view of galaxies is bound to affect estimates of their properties. For instance, when estimating nebular metallicities, one would in principle need to count only the emission-line photons produced in star-forming (SF) regions, isolating them from those coming from other nebular regimes, like the diffuse ionized gas (DIG), or photoionization by an active nucleus or old stars. Similarly, estimates of the dust attenuation and star formation rate from a single spectrum forcefully neglect variations in dust content across the face of galaxies. A proper assessment of these effects is important to better understand the precise meaning of (and potential biases in) properties derived from integrated-light data.

As a starting step towards this goal, this paper takes advantage of integral field spectroscopy from the CALIFA survey \citep{SFSanchez.DR3.2016} to study the importance and characteristics of the DIG in different regions of galaxies of all types. The complete coverage in terms of Hubble types and inclinations makes CALIFA ideal for a first general exploration of the DIG in galaxies of the local Universe, even though its spatial resolution does not allow for a detailed description of the different interstellar medium (ISM) components.  Imaging spectroscopy of galaxies with higher spatial resolution is developing \citep[e.g.,][]{Sanchez.etal.2015MUSE, Vogt.etal.2017a, RousseauNepton.etal.2017} but the data available so far cover  too few objects for an overall study as the one presented here.

The DIG was first detected in the Galactic disc through faint emission-lines outside the classical \hii regions \citep{Reynolds.PhD.1971}. Observations of edge-on spiral galaxies using deep \Ha imaging \citep{Dettmar.1990, HoopesWaltGreen.1996, HoopesWaltRand.1999} showed the existence of DIG not only in the vicinity of \hii regions but also at large distances above galaxy planes. By studying 109 \hi selected galaxies in the SINGS survey, \citet{Oey.etal.2007} came to the conclusion that diffuse \Ha emission is present in galaxies of all types and represents about 60 per cent of the total \Ha emission, irrespective of the galaxy Hubble type or total star formation rate.

Radiation from massive OB stars leaking out from \hii regions is a commonly advocated ionization source for the DIG \citep[see review by][]{Haffner.etal.2009}. However, the existence of additional/alternative sources is required by the increase  of \nii/\Ha, \sii/\Ha, and \oiii/\Hb with galactic height found in many galaxies.

Such features cannot be reproduced with models of photoionization by hot, massive stars, even taking into account the hardening of the ionizing radiation due to intervening absorption \citep{HoopesWalt.2003}. The most commonly invoked sources of additional ionization/heating  are shocks \citep{CollinsRand.2001}, turbulent mixing layers \citep{SlavinShullBegelman.1993, Binette.etal.2009a}, magnetic reconnection, cosmic rays or photoelectric emission from small grains \citep{Reynolds.etal.2001}, and hot low-mass evolved stars \citep[HOLMES;][]{FloresFajardo.etal.2011a}. HOLMES have also been invoked as the ionization source of the weak emission-lines in retired galaxies (\citealt{Stasinska.etal.2008a} and \citealt[hereafter CF11]{CidFernandes.etal.2011a}). These are systems that have stopped forming stars and are ionized by their hot, old stellar populations, producing LINER-like  emission-line ratios, a phenomenon that is common in both ellipticals and in bulges of spiral galaxies \citep{Sarzi.etal.2010, Gomes.etal.2016a, Belfiore.etal.2016}.

Regardless of what powers the DIG, its nebular regime differs from that in  \hii regions, with lower densities, lower ionization parameters, and higher electron temperatures. Failure to account for its contribution may thus lead to biases in the derived properties of galaxies.

In the literature, DIG and SF regions are separated on the basis of the \Ha surface brightness, $\Sigma_{\Ha}$. This is a natural criterion, since the \Ha surface brightness is directly related to the density of the ionized gas. \cite{Zhang.etal.2017a}, for instance, use $\Sigma_{\Ha} > 10^{39}\,$erg$\,$s$^{-1}\,$kpc$^{-2}$ to `select reliable \hii-dominated spaxels'. In other studies, the criterion is not a simple $\Sigma_{\Ha}$ threshold, but still based on $\Sigma_{\Ha}$ (see discussions in \citealt{Zurita.etal.2000}, \citealt{Oey.etal.2007}, and \citealt{Vogt.etal.2017a}). This intuitively valid approach is however not fully adequate. As argued in this paper, separating SF and DIG regions on the basis of $\Sigma_{\Ha}$ is conceptually incorrect, and may lead to inconsistent results under certain circumstances. Furthermore, $\Sigma_{\Ha}$ gives no clue as to the nature of the DIG emission.

These drawbacks are solved with a simple diagnostic based on the equivalent width of \Ha ($W_{\Ha}$). As shown in this paper, $W_{\Ha}$ correctly tracks the qualitative differences inherent to the SF and DIG regimes, and is also capable of identifying the component of the DIG corresponding to gas predominantly ionized by HOLMES, the hDIG, in an observationally simple and physically sound fashion. The nature of the remaining (neither HOLMES nor SF-dominated) DIG emission is most probably a mixture of processes, and will hereafter be dubbed mDIG, where the `m' stands for `mixed'.

This paper is organized as follows. Section \ref{sec:Data} describes the data and processing steps. Section \ref{sec:DIGxHII} presents our $W_{\Ha}$-based method to separate DIG from SF spaxels, and to distinguish the mDIG and hDIG components. Section \ref{sec:Discussion} applies the method to CALIFA data to study hDIG/mDIG/SF fractions and the nature of extraplanar DIG emission in edge-on systems. Comparisons with  $\Sigma_{\Ha}$-based methods and a classical excitation diagram analysis are also presented. Finally, Section \ref{sec:Conclusions} highlights our main findings.

\section{Data}
\label{sec:Data}

The spatially resolved 3650--6850 \AA\ spectra covering the whole optical extent of relatively nearby galaxies gathered by CALIFA \citep{Sanchez.etal.2012a, SFSanchez.DR3.2016, Husemann.etal.2013a, GarciaBenito.etal.2015a} provide a suitable data set to address the issues raised above. We use the COMBO data cubes, obtained by merging the observations with the two grisms used by CALIFA. The spectral resolution is 6 \AA\ in full width at half-maximum, the field of view is slightly over 1 arcmin${}^2$, and the spaxel size is $1 \times 1$ arcsec$^2$, but the spatial resolution is $\sim 2.5$ arcsec. At the distance of our sources (20--123 Mpc), this corresponds to  0.2--1.5 kpc (0.8 kpc on the median).

Our working sample contains 391 galaxies, morphologically distributed as follows: 57 ellipticals, 47 S0--S0a, 62 Sa--Sab, 67 Sb, 70 Sbc, and 88 Sc or later. These same six bins in galaxy morphology will be used to evaluate how the SF, mDIG, and hDIG components vary across the Hubble sequence. Morphologically distorted systems (such as those studied by \citealt{Wild.etal.2014, BB.etal.2015b, BB.etal.2015a, CortijoFerrero.etal.2017b, CortijoFerrero.etal.2017a}) were discarded from the analysis. Besides its diversity in Hubble types, the sample also covers inclination angles from edge-on to face-on.

All data cubes were processed though the \pycasso\footnote{Python CALIFA \starlight\ Synthesis Organizer} pipeline described in \citet{CidFernandes.etal.2013a} and \citet{Andre.etal.2017}. Briefly, after masking artefacts, foreground sources, and low signal-to-noise ($SN$) regions, the data cubes are binned into Voronoi zones gauged to reach an $SN$ of 20 or more in the continuum around 5635 \AA. Our sample contains 307958 zones ($\sim 800$ per galaxy). These zone spectra are then processed through the {\sc starlight} code \citep{CidFernandes.etal.2005a}, obtaining a model $M_\lambda$ for the stellar continuum. Previous papers using this pipeline concentrated on the analysis of the spatially resolved properties of the stellar populations as deduced from the spectral fits \citep{Perez.etal.2013, GonzalezDelgado.etal.2014b, GonzalezDelgado.etal.2015a, GonzalezDelgado.etal.2016a, GonzalezDelgado.etal.2017}.  Instead, this paper focuses on emission-line properties.

Emission-line fluxes are measured with the {\sc sherpa} IFU line fitting software \citep[SHIFU;][]{RGB.etal.2017}, based on {\sc ciao's sherpa} package \citep{Freeman.etal.2001, Doe.etal.2007} by fitting single Gaussians to the $R_\lambda = O_\lambda - M_\lambda$ residual spectra obtained after subtracting the \starlight\ fit ($M_\lambda$) from the observed spectrum ($O_\lambda$). These can be delicate measurements in the cases where lines are weak, as it is often the case with \Hb. This study deals almost exclusively with \Ha, whose flux is much less affected by uncertainties. Indeed, the median $SN_{\Ha}$ is 16 over all our zones, and in only 5 per cent of the cases $SN_{\Ha} < 1$.

\begin{figure*}
\includegraphics{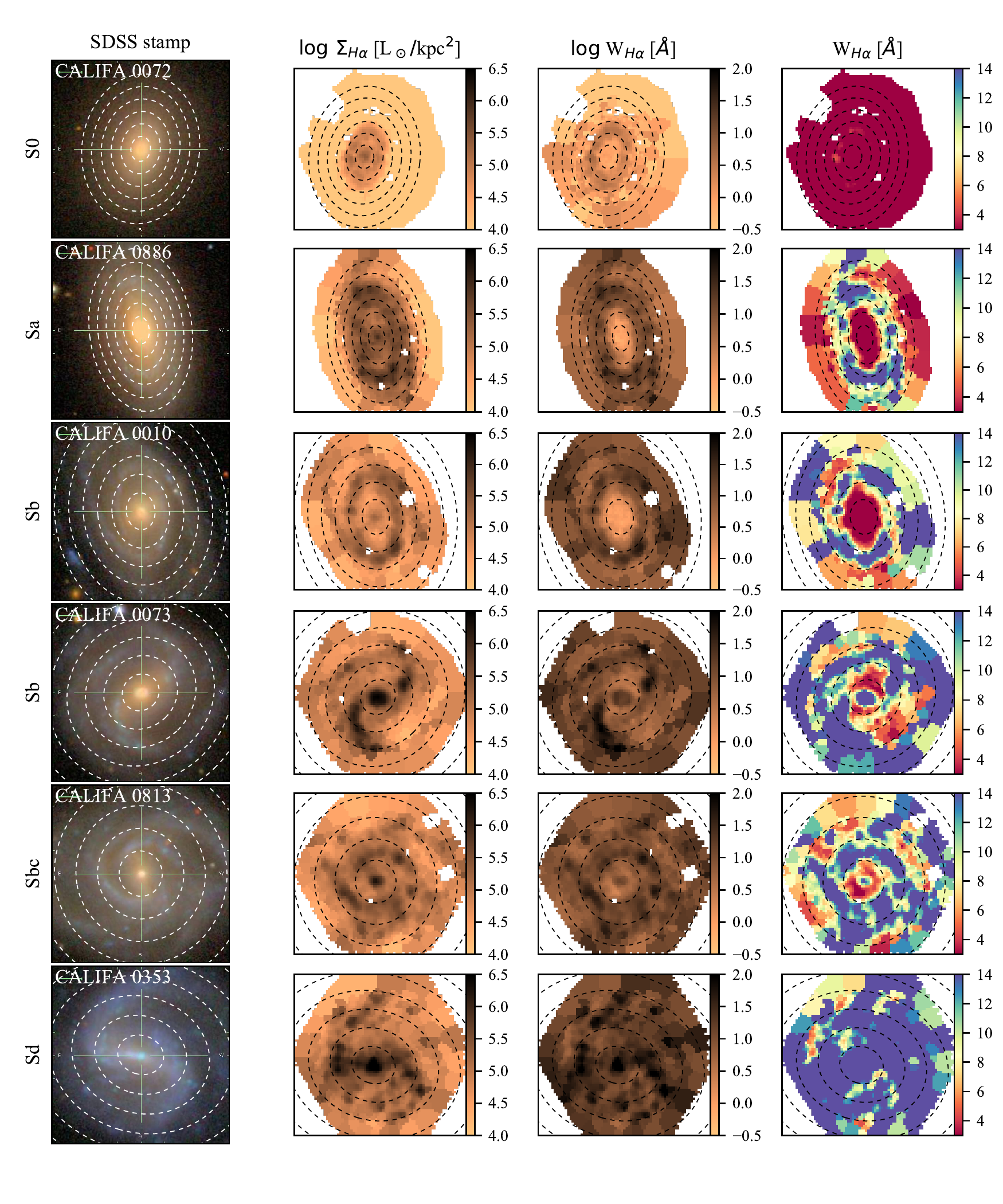}
 \caption{SDSS stamps, $\Sigma_{\Ha}$, and $W_{\Ha}$ maps for example galaxies in CALIFA. Images on the right show $W_{\Ha}$ maps saturated at 3 and 14 \AA, highlighting the proposed classification of hDIG, mDIG, and SFc. Dashed elliptical rings mark radial distances to the nucleus of $R = 0.5$, 1.0, 1.5, \ldots in units of the galaxy's half-light radius (HLR). Empty patches mask foreground sources and other artefacts.
 }
 \label{fig:ExampleMaps}
\end{figure*}

Fig.\ \ref{fig:ExampleMaps} shows SDSS stamps along with our CALIFA-based $\Sigma_{\Ha}$ and $W_{\Ha}$ maps for a selection of galaxies in our sample. Dotted ellipses mark the distance to the nucleus ($R$), measured along the major axis of ellipses defined on the basis of the spatial moments of the continuum flux in the (rest-frame) $5635 \pm 45$ \AA\ continuum. We measure $R$ in units of the half-light radius (HLR), measured in the same continuum window around $\lambda = 5635$ \AA. As illustrated by the studies of \citet{Perez.etal.2013}, \citet{Sanchez.etal.2014}, and \citet{GonzalezDelgado.etal.2016a}, the HLR is a convenient metric for comparing galaxies of different sizes. For our galaxies, HLR $= 3.9 \pm 1.7$ kpc (mean $\pm$ dispersion). In spirals, one may generally associate $R > 1$ HLR with the disc and $R < 0.5$ HLR with the bulge. The meaning of $R$ becomes ambiguous for highly inclined systems, a limitation that does not affect our main results.

%---------------------------- Figure ----------------------------
%---------------------------- Figure ----------------------------

The constant patches in the example $\Sigma_{\Ha}$ and $W_{\Ha}$ maps in Fig.\ \ref{fig:ExampleMaps} correspond to the Voronoi zones used to ensure reliable continuum spectra to be processed through {\sc starlight}. As seen in the figure, these are always located in the fainter, outer parts of our galaxies. Within $R < 1$, HLR over 97 per cent of the individual spaxels have $SN > 20$, and so no spatial binning is performed. Of the 307958 spectra in our sample, 274534 (89 per cent) correspond to single spaxels. The remaining ones correspond to Voronoi zones containing six spaxels in the median.

Part of the analysis that follows is based on the statistics of $W_{\Ha}$ among the spectra in our sample. The variable size of our extractions introduces some distortion in these statistics, since a large zone counts the same as one containing a single spaxel. The effects of this distortion will be discussed below, but we anticipate that they do not affect the main results reported in this paper.

\section{$W_{\Ha}$ as a diagnostic of the nebular regime: SF\lowercase{c}, \lowercase{m}DIG, and \lowercase{h}DIG}
\label{sec:DIGxHII}

We wish to devise a method to characterize the dominant nebular regime in the various galaxy regions, one that is able to distinguish SF- from DIG-dominated zones, and also to differentiate  DIG components. The ulterior goal of establishing such a taxonomical scheme is to provide a basis for other studies. With a solid classification scheme, one can, for instance, perform a comparative study of the physical properties (say, dust content or stellar populations) associated with these components of the ISM, or evaluate possible biases resulting from the mix of different nebular regimes in single-fibre (one spectrum per galaxy) observations.

Given that our spatial resolution of  $\sim 0.8$ kpc exceeds the physical scale of \hii regions (even giant ones like 30 Dor or NGC 604, which are typically 0.1--0.3 kpc across, e.g.; \citealt{Rosa.y.Enrique.2000}), SF regions in our data are bound to contain diffuse gas emission. Naturally, this mixture is even stronger in regions spatially binned into Voronoi zones. To convey this relevant fact in our nomenclature, we will hereafter use the term star-forming complexes (SFc) to refer to regions that contain \hii regions, but inevitably mixed with DIG emission in our data. A more precise definition is that our SFc are the zones with a larger SF/DIG ratio. Note that our SFc are not even necessarily \textit{dominated} by star formation, but simply contain a good proportion of SF-powered line emission.  This relative, ranking-based scale is implicit in all that follows.

\subsection{Rationale for a $W_{\Ha}$-based SF/DIG separator}
\label{sec:WHaRationale}

Previous work favours the use of the \Ha\ surface brightness ($\Sigma_{\Ha}$) as a means to separate the DIG from SF regions. For instance, in a recent study, \cite{Zhang.etal.2017a} claim that, for MaNGA \cite{Bundy.etal.2015} data, a $\Sigma_{\Ha} > \Sigma_{\Ha}^{\rm SF,min} = 10^{39}$ erg$\,$s$^{-1}\,$kpc$^{-2}$ criterion selects reliable SF-dominated spaxels.

We prefer to distinguish SF from DIG  on the basis of the equivalent width of \Ha. A simple thought experiment suffices to demonstrate that $W_{\Ha}$ offers a more suitable way to distinguish DIG from SF regions than $\Sigma_{\Ha}$.

Consider combining the emission of two identical DIG-dominated cubic volume elements, each of area  $A$ and emitting a flux $F_{\Ha} = A \times \Sigma_{\Ha}$. As is appropriate for diffuse gas, suppose also the medium is optically thin to the \Ha photons, so that one sees the whole volume. Clearly, this DIG + DIG operation should not alter the identification of the nebular regime in the combined element, which should still be identified as DIG. Seen side by side, the joint surface brightness of the two elements will also be $\Sigma_{\Ha}$ (twice the flux over twice the area). If, however, the two cubes were to be placed along the same line of sight, one would see twice the \Ha flux over the same area $A$, and hence a doubling of the surface brightness. A $\Sigma_{\Ha}$-based criterion will thus lead to a DIG $+$ DIG $=$ SF logical inconsistency if the original elements are brighter than half the chosen $\Sigma_{\Ha}^{\rm SF, min}$ threshold. A $W_{\Ha}$-based criterion, on the other hand, would lead to a consistent DIG classification independent of how the merged element is seen, as the final equivalent width is the same as that of the original elements.

As shown in Section \ref{sec:SBHa}, the difference between $\Sigma_{\Ha}$- and $W_{\Ha}$-based criteria is particularly important in bulges, whose geometrically long path lengths may lead to $\Sigma_{\Ha} > \Sigma_{\Ha}^{\rm SF, min}$  even in the absence of SF.

A further and independent argument in favour of a $W_{\Ha}$-based criterion is that properties such as colours, stellar mass density, gas content, and others show a radial dependence in galaxies, so a criterion based on a fixed threshold is likely not  appropriate for all the regions of a galaxy. In particular, in the case  where the DIG emission is powered by HOLMES, the approximately constant ratio of ionizing photons per unit mass of old stellar populations implies a $W_{\Ha}$ of the order of 1 \AA\ (\citealt{Binette.etal.1994a}; CF11; \citealt{Belfiore.etal.2016}), independently of the total line fluxes involved, so that one may well have \Ha-bright HOLMES-powered regions (the hDIG) mistakenly classified as SF with a $\Sigma_{\Ha}$-based criterion. Conversely, faint SF regions may be misclassified as DIG because of a low $\Sigma_{\Ha}$.

In both arguments, it is the extensive nature of $\Sigma_{\Ha}$ that makes it prone to confusing DIG and SF regions. $W_{\Ha}$, in contrast, is an intensive property\footnote{The qualifications  `intensive' and `extensive' are used here in analogy with their thermodynamic connotation: intensive properties do not depend on the size or mass of a system (the column projected in a spaxel, in our case), whereas extensive properties are additive. The juxtaposition, either in 3D or just in projection, of two identical volume elements, each emitting a line flux $F$ and a continuum flux density $C$ over an area $A$, results in a $W = 2F/2C$ equivalent width identical to that of the individual elements, so that $W$ is an intensive property. On the other hand, one now has twice the flux exiting the same area, so that $\Sigma = 2F/A$ equals the sum of the individual $\Sigma$'s. In this sense, surface brightness behaves as an extensive property.}, and one that passes the two consistency tests posed above.

\subsection{The observed distribution of $W_{\Ha}$ and the hDIG component}
\label{sec:hDIG}

%---------------------------- Figure ----------------------------
\begin{figure*}
\includegraphics{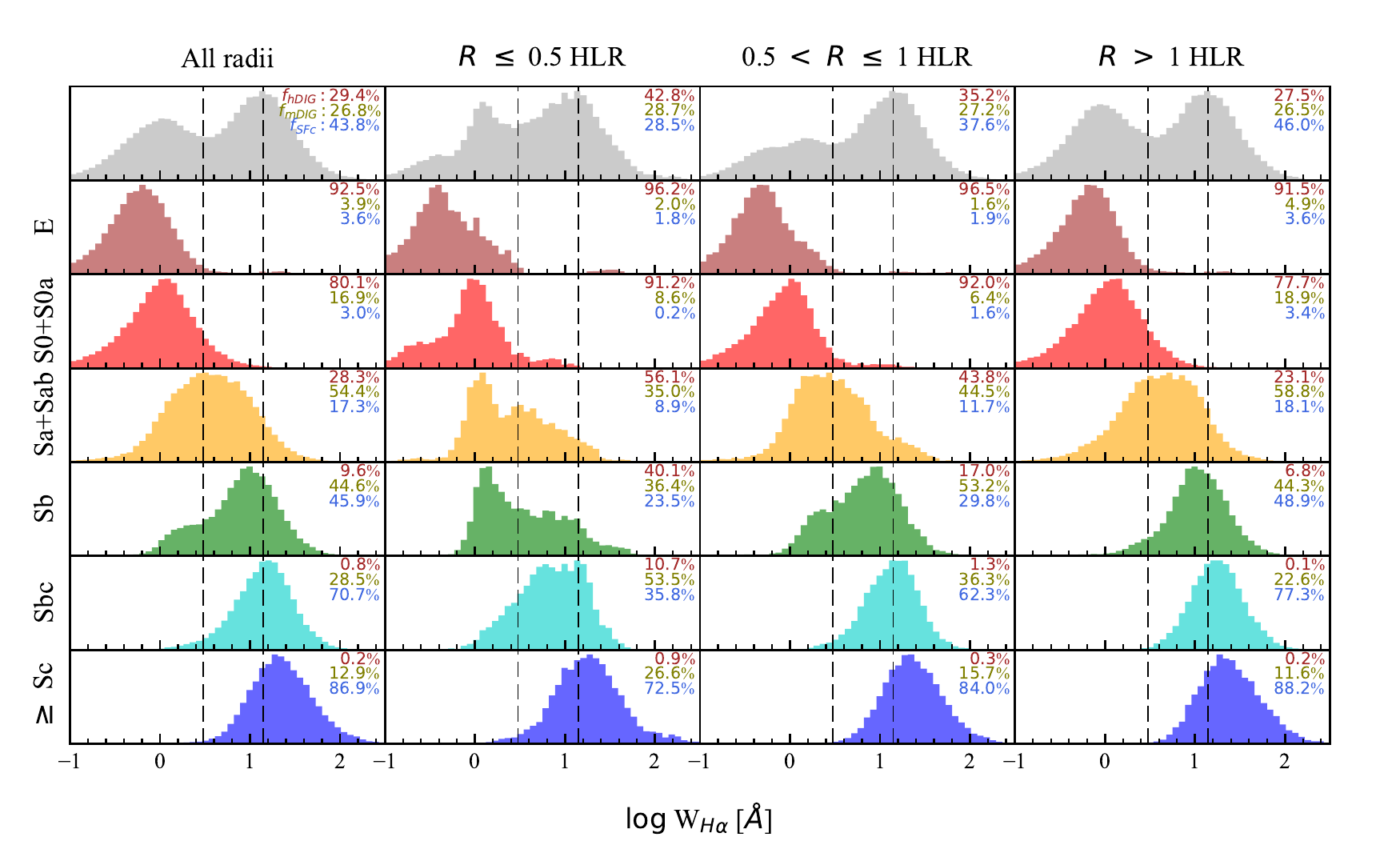}
\caption{Distribution of $W_{\Ha}$ among $307\,958$ zones of 391 CALIFA galaxies. Different rows show the breakdown of this distribution by Hubble type, from ellipticals (second row) to Sc and later (bottom). Results for the whole sample are shown in the top row. Histograms in the left-hand panels count all zones, while the others select different ranges in radius: the inner 0.5 HLR (second column), $R = 0.5$--1 HLR (third), and outwards of 1 HLR (fourth). Vertical dashed lines mark the hDIG/mDIG and mDIG/SF frontiers at 3 and 14 \AA\, respectively. Labels on the top right of each panel list the fraction of the \Ha flux associated with the hDIG, mDIG, and SF components (averaged over galaxies in each panel).
}
 \label{fig:WHaDistrib_ALLgals}
\end{figure*}
%---------------------------- Figure ----------------------------

Having made the case for a $W_{\Ha}$-based characterization of DIG regions, this section presents empirical evidence that guides the implementation of $W_{\Ha}$ thresholds to separate SF from DIG, and to identify the HOLMES-dominated component.

The observed distribution of $W_{\Ha}$ among and within galaxies offers valuable insight on this issue. Fig.\ \ref{fig:WHaDistrib_ALLgals} shows $W_{\Ha}$ histograms for $\sim 300$k zones from the 391 CALIFA data cubes in our sample. The top panels show the result for the full sample, while the other rows split the sample by morphological types, from ellipticals in the second row to Sc and later in the bottom. Histograms in the left column count all zones, whereas columns to the right show results for different radial regions: $R < 0.5$, 0.5--1, and $> 1$ HLR.

The global distribution (top-left panel) is strongly bimodal, with a low $W_{\Ha}$-population peaking at $W_{\Ha} \sim 1$ \AA\ and a higher one at $\sim 14$ \AA. This behaviour is remarkably similar to that seen in SDSS galaxies (\citealt{Bamford.etal.2008a}; CF11).Previous spatially resolved studies based on both CALIFA \citep{Morisset.etal.2016} and MaNGA data \citep{Belfiore.etal.2016, Belfiore.etal.2017} have also identified this bimodality in $W_{\Ha}$.

The relative amplitudes of the two peaks in the $W_{\Ha}$ distribution are sensitive to the spatial binning scheme. Without Voronoi binning, and restricting the analysis to the inner 2 HLR to eliminate the noisiest spectra, we find that the high-$W_{\Ha}$ population at $R > 1$ HLR increases by about a factor of 3 with respect to that seen in the top right panel of Fig.\ \ref{fig:WHaDistrib_ALLgals}. This increase comes almost exclusively from galaxies of Hubble type Sb or later. The low-$W_{\Ha}$ peak, on the other hand, changes by just  $\sim 20$ per cent. The bimodality, however, is preserved. Indeed, two-Gaussians fits to the pixel and zone-based distributions identify similar components, with peaks centred at $\sim 1$ and 14 \AA\ in both cases.

\begin{figure}
 \includegraphics{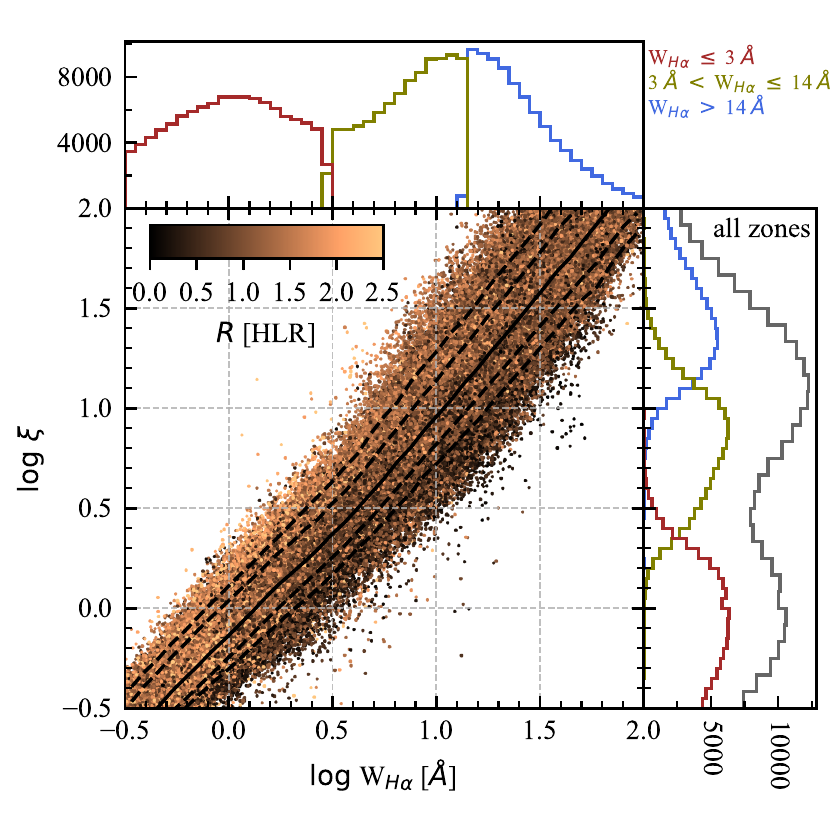}
 \caption{ Ratio  between the observed \Ha luminosity and that predicted from populations older than $10^8$ yr ($\xi$) as a function of  $W_{\Ha}$. Points are colour-coded by the radial distance to the nucleus (in units of HLR). The histograms of $\xi$ and $W_{\Ha}$  are colour-coded as hDIG/mDIG/SFc (red/yellow/blue), showing that low-$W_{\Ha}$ regions are compatible with ionization by HOLMES ($\xi \sim 1$). }
 \label{fig:WHa-Xi}
\end{figure}

We interpret the low-$W_{\Ha}$ population as representing gas photoionized by HOLMES. To test this, we followed the methodology in CF11 by computing the ratio $\xi$ between the observed \Ha luminosity and that predicted from the ionizing photons produced by populations older than $10^8$ yr (inferred from the {\sc starlight} analysis). Since the stellar population models used in our \starlight\ fits \citep{GonzalezDelgado2005,Vazdekis2010} do not extend to $h\nu > 13.6$ eV, we have borrowed the number of ionizing photons from \citet{Bruzual.Charlot.2003} for a Salpeter initial mass function and \cite{Girardi2000} tracks. As discussed in CF11, different sets of models lead to systematic differences of 0.2--0.5 dex in the predicted ionizing fluxes. Fig.~\ref{fig:WHa-Xi} shows $\xi$ as a function of  $W_{\Ha}$, with histograms colour-coded by our hDIG/mDIG/SFc classification. We find that $\xi$ is indeed of the order of 1 for zones in the low-$W_{\Ha}$ peak. Hence, despite all uncertainties involved in this computation (CF11; \citealt{Belfiore.etal.2016, Morisset.etal.2016}), the end result corroborates the interpretation that HOLMES are responsible for the low-$W_{\Ha}$ population.

%---------------------------- Figure ----------------------------
%---------------------------- Figure ----------------------------

The correspondence of this interpretation with the concept of retired galaxy put forth by \citet{Stasinska.etal.2008a} is evident. These are systems that have stopped forming stars long ago and whose ionizing photon budget is dominated by hot post-asymptotic giant branch stars and white dwarfs, leading to $W_{\Ha}$ values of the order of $\sim 1$ \AA. Furthermore, the minimum seen at $W_{\Ha} \sim 3$ \AA\ coincides with the threshold proposed by CF11 to distinguish retired galaxies from those where SF or AGN activity dominates the line emission. We thus claim that $W_{\Ha} < 3$ \AA\ regions should be treated as HOLMES-ionized gas, the hDIG, a sub-type of DIG emission.

The breakdown of the $W_{\Ha}$ distribution by Hubble type in Fig.\ \ref{fig:WHaDistrib_ALLgals} suggests that the bimodality is always present, but the proportion of the low- to high-$W_{\Ha}$ populations shifts with morphology: early-type galaxies are overwhelmingly dominated by values around the $\sim 1$ \AA\ peak, well within the hDIG regime, while in late-type spirals it is the higher $W_{\Ha}$  population that dominates.

The radial dependence of the $W_{\Ha}$ distribution offers further insight. The second to last columns in Fig.\ \ref{fig:WHaDistrib_ALLgals} show that the hDIG population is evenly spread in $R$ for early-type galaxies, confirming earlier CALIFA-based studies by \citet{Kehrig.etal.2012}, \citet{Singh.etal.2013}, and \citet{Gomes.etal.2016b}, as well as the MaNGA-based analysis by \citet{Belfiore.etal.2016, Belfiore.etal.2017}. Among Sb and later type spirals, on the other hand, the hDIG is evidently concentrated in the central regions of galaxies. To put this in numbers, 82 per cent of the $W_{\Ha} < 3$ \AA\ points in the 225 Sb or later type galaxies are located within $R < 1$ HLR.

We interpret this higher incidence of hDIG zones in the central regions of galaxies as a corollary of the prevalence of old stellar populations in galaxy bulges. The ionizing photon budget in these retired bulges is dominated by HOLMES, as any relevant contribution from other sources would raise their observed $W_{\Ha}$ to larger values.

Conversely, the low incidence of $W_{\Ha} < 3$ \AA\ zones at large $R$ in spirals indicates that pure hDIG emission is not a statistically relevant component of the DIG that permeates the space between SF regions in galaxy disks. This is not meant as a general conclusion, as Fig.\  \ref{fig:WHaDistrib_ALLgals} itself suggests that HOLMES may explain a substantial part of the disc emission in Sa--Sab galaxies. Among Sb and later type systems, however, hDIG-dominated disc zones are rare.

The motivation to introduce the hDIG category is thus firmly rooted on both observational and theoretical arguments. This well-understood component of the DIG becomes dominant whenever old stellar populations are the most relevant source of ionizing photons.

We close this section by noting that experiments were carried out to investigate to which extent galaxy inclination affects the $W_{\Ha}$ distributions depicted in Fig.\  \ref{fig:WHaDistrib_ALLgals}. This was done by first eliminating E and S0's and then splitting the sample into bins in minor-to-major axis ratio $b/a$ (measured as explained in \citealt{Andre.etal.2017}). No major effect is found. The only noteworthy trend identified is that for zones within $R < 0.5$ HLR, the histograms tend to shift towards smaller $W_{\Ha}$ (by $\sim 0.2$--0.3 dex on the median) as one goes from edge-on to face-on viewing angles, a tendency that is understood as a simple projection effect. While face-on views of these inner regions sample the bulge, with its characteristic hDIG emission, as the inclination increases, parts of the disc and get projected upon the bulge, resulting in a mixture of  hDIG and SFc zones. Thus, as for the Voronoi binning effects discussed above, inclination effects do  not erase the fundamental dichotomy between these two nebular regimes.

\subsection{Identification of hDIG, mDIG, and SFc components}
\label{sec:WHaLimits}

Unlike the low-$W_{\Ha}$ regions, which can be safely associated with an hDIG regime, zones belonging to the high-$W_{\Ha}$ population cannot be unequivocally tagged as SFc. Sure enough, SFc are among those with $W_{\Ha} > 3$ \AA, but this population includes other processes too. In particular, diffuse gas powered by ionizing radiation leaking from \hii regions is also part of this population, with a ratio of ionizing photons per unit stellar mass leading to $W_{\Ha}$ values above those typical of the HOLMES-dominated regime.

Though the whole $W_{\Ha} > 3$ \AA\ population ultimately comprises a mixture of regimes, it is useful to sub-divide it into mDIG and SFc classes in order to identify zones where SF is comparatively more important. There is no conspicuous boundary that cleanly separates SFc from mDIG in terms of $W_{\Ha}$, however. Indeed, the continuous, unimodal behaviour of the $W_{\Ha} > 3$ \AA\ population in Fig.\  \ref{fig:WHaDistrib_ALLgals} is not suggestive of sub-populations, but of a continuous distribution. In the absence of a clear-cut criterion, we place the mDIG/SFc division at $W_{\Ha} = 14$ \AA, coinciding with the peak of the high $W_{\Ha}$ population in the histograms in Fig.\  \ref{fig:WHaDistrib_ALLgals}.

Our final hDIG/mDIG/SFc classification scheme thus becomes

\begin{itemize}
 \item hDIG: $W_{\Ha} < 3 \,\mathrm{\AA}$,
 \item mDIG:  $3 \,\mathrm{\AA}  < W_{\Ha} < 14 \,\mathrm{\AA}$,
 \item SFc: $W_{\Ha} > 14 \,\mathrm{\AA}$.
\end{itemize}

The reader should take note of a marked conceptual asymmetry in these definitions. While the mDIG/hDIG frontier at 3 \AA\ is based on a firm theoretical understanding of the nature of the hDIG population, fully corroborated by the bimodal distribution of $W_{\Ha}$, nothing of the sort can be said about the mDIG/SFc division. All that can be said about regions above the $W_{\Ha} = 14$ \AA\ limit is that they have a higher proportion of SFc than those below it. This scheme should thus be used with the understanding that our mDIG regions may well host some star formation, and that  $W_{\Ha} > 14$ \AA\ does not isolate  pure SF regions. Bona fide giant \hii regions, those that are the basis of any emission-line study of galaxies, have $W_{\Ha}$ an order of magnitude larger \citep{McCall.etal.1985, Garnett.and.Shields.1987, Kennicutt.and.Garnett.1996, Luridiana.and.Peimbert.2001, Bresolin.etal.2004}, but these are heavily diluted at our resolution.

The rightmost panels in Fig.\ \ref{fig:ExampleMaps} show $W_{\Ha}$ maps with a colour scheme designed to saturate at $< 3$ and $> 14$ \AA. hDIG regions are thus depicted in red and SFc ones in blue, with intermediate colours used to trace the 3--14 \AA\  mDIG range. The S0 galaxy at the top of the plot exemplifies the dominance of the hDIG component amongst early-type galaxies, as previously inferred from their $W_{\Ha}$ histograms in Fig.~\ref{fig:WHaDistrib_ALLgals}. The  prevalence of this same component in bulges is also illustrated in this figure, particularly in the cases of CALIFA 0886 (NGC 7311) and 0010 (NGC 0036). As expected, SFc become increasingly important as one moves down the Hubble sequence (top to bottom in  Figs.\ \ref{fig:ExampleMaps} and \ref{fig:WHaDistrib_ALLgals}).

The reader is referred to \citet{Sanchez.etal.2015MUSE} for an example of a $W_{\Ha}$ map obtained under much higher spatial resolution. Their MUSE-image of the spiral galaxy NGC 6754 shows a plethora of SFc, embedded in a smoother medium with mDIG-like $W_{\Ha}$ values that permeates the whole disc.

\section{Discussion}
\label{sec:Discussion}

The theoretically and empirically inspired set of criteria to identify hDIG, mDIG, and SFc in galaxies serves a variety of purposes. In this section, we apply them to our CALIFA data with the specific goals of \textit{(i)} estimating the relative strengths of hDIG, mDIG, and SFc components in galaxies across the Hubble sequence (Section \ref{sec:CurveOfGrowth}), \textit{(ii)} studying the nature of extraplanar diffuse line emission in edge-on systems (Section \ref{sec:EdgeOn}), \textit{(iii)} comparing results obtained with our method with those derived with a $\Sigma_{\Ha}$-based SF/DIG separation scheme (Section \ref{sec:SBHa}), \textit{(iv)} investigating the possibility of differentiating SF and DIG regimes with density-sensitive line ratios (Section \ref{sec:nSii}), \textit{(v)} testing the consistency of our criteria with a classical diagnostic diagram analysis (Section \ref{sec:BPT}), and \textit{(vi)} investigating the mDIG mixture (Section \ref{sec:TheNatureOfTheMIG}). We close with a discussion of caveats involved (Section \ref{sec:Caveats}).

\subsection{The relative strengths of the hDIG, mDIG, and SFc components}
\label{sec:CurveOfGrowth}

%---------------------------- Figure ----------------------------
\begin{figure}
 \includegraphics{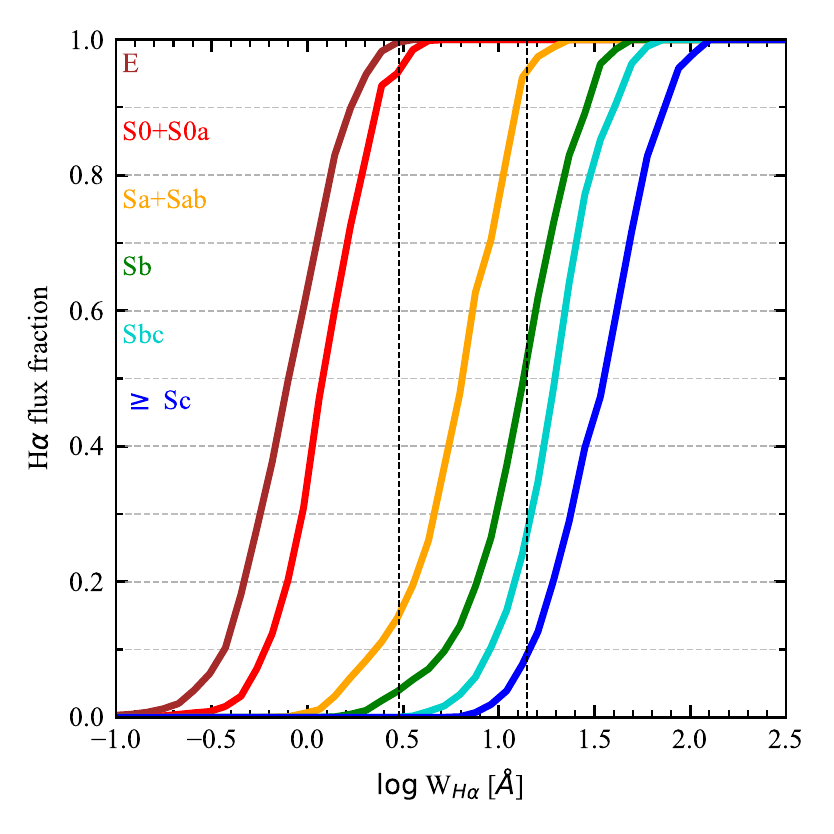}
 \caption{Cumulative fraction of the total galaxy \Ha flux coming from regions with $W_{\Ha}$ smaller than a given value. The plot shows the median curves obtained for galaxies in six Hubble types.}
 \label{fig:CurveOfGrowth}
\end{figure}
%---------------------------- Figure ----------------------------

One of the questions that can be addressed with the classification proposed above is what are the relative strengths of the hDIG, mDIG, and SFc components, and how these proportions vary as a function of Hubble type. This issue bears on the interpretation of properties derived through spatially unresolved spectroscopic data, where these components cannot be separated, like for galaxies at high redshifts, for example.

A simple and observationally relevant  way to quantify this is to compute the fractional contribution of each component to the total \Ha flux of a galaxy. For the galaxies in Fig.\ \ref{fig:ExampleMaps}, for instance, these fractions range from $(f_{\rm hDIG} , f_{\rm mDIG} , f_{\rm SFc}) = (87,13,0)$ for the S0 galaxy CALIFA 0072, to $(5.5,47,47.5)$ for the Sb galaxy CALIFA 0010, and $(0.3,46.1,52.6)$ for CALIFA 0813, an Sbc. This steady progression along the Hubble sequence reflects the tendencies seen in Fig.\ \ref{fig:WHaDistrib_ALLgals}, where the values of $(f_{\rm hDIG} , f_{\rm mDIG} , f_{\rm SFc})$ are given in each panel, along with the $W_{\Ha}$ histogram, for different radial regions and morphological types.

Fig.\ \ref{fig:CurveOfGrowth} presents these fractions for the whole sample in a more elaborate way. For each galaxy, we compute the cumulative fraction ($f$) of the total \Ha flux coming from zones with $W_{\Ha}$ smaller than a given value. The resulting $f(<W_{\Ha})$ growth curves yield not only the $(f_{\rm hDIG} , f_{\rm mDIG} , f_{\rm SFc})$ fractions, but a more continuous depiction of the makeup of a galaxy's \Ha output in terms of $W_{\Ha}$. The figure shows the median curves obtained for each of our six bins in Hubble type. Vertical dashed lines mark the hDIG/mDIG and mDIG/SFc frontiers at 3 and 14 \AA\, respectively.

The steady progression from early to late types confirms the expectation from their $W_{\Ha}$ distributions  (Fig.\ \ref{fig:WHaDistrib_ALLgals}), and allows us to quantify the relative importance of each component to the total \Ha flux. Ellipticals and S0's have nearly all of their \Ha in the hDIG phase ($W_{\Ha} < 3$ \AA). Among Sa--Sab systems, this component accounts for $\sim 14 per cent$ of \Ha, with mDIG emission contributing most of the remaining flux. From Sb onwards, the SFc fraction is 50 per cent or larger. Note that these fractions reflect the median behaviour. Naturally, there is substantial scatter from galaxy to galaxy even for a fixed morphological class.

DIG fractional contribution to \Ha fluxes has been estimated in several previous studies based on narrow-band (\Ha+\ \nii) imaging \citep{Ferguson.etal.1996, Zurita.etal.2000, Thilker.etal.2002, Oey.etal.2007}, with results varying substantially due to differences in the methodology to separate the DIG emission. The largest study to date is that by \citet{Oey.etal.2007}, who estimate a mean DIG fraction of $59\pm19 per cent$ among 109 galaxies in the SINGG survey \citep{Meurer.etal.2006}. For our sample and definitions, we find a very similar DIG (mDIG + hDIG) fraction of 56 per cent, but a larger dispersion ($\pm 38$ per cent). Unlike in our study (Fig.\ \ref{fig:CurveOfGrowth}), they find no evidence for a correlation with Hubble type, a difference that may be due to sample selection criteria and methodology to define DIG/SF.

\subsection{Edge-on systems: the nature of extraplanar line emission}
\label{sec:EdgeOn}

%---------------------------- Figure ----------------------------
\begin{figure*}
 \includegraphics{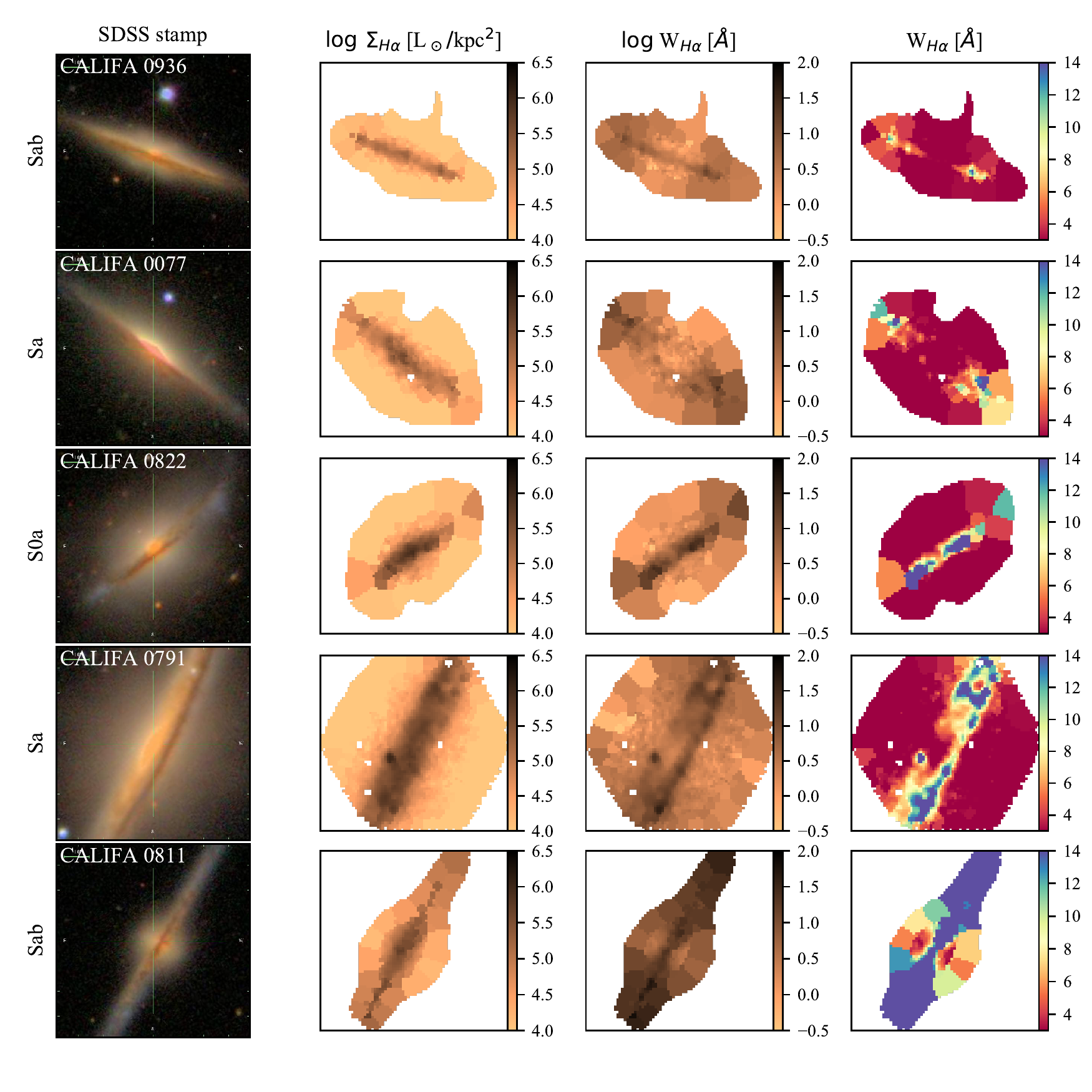}
 \caption{As Fig.\ \ref{fig:ExampleMaps}, but for edge-on galaxies.}
 \label{fig:ExampleMapsEdgeOn}
\end{figure*}
%---------------------------- Figure ----------------------------

Edge-on galaxies are important for the study of the DIG, as they exhibit a systematic behaviour of emission-line properties with height above the galactic discs \citep{Tullmann.and.Dettmar.2000, Otte.etal.2002, Jones.etal.2017}. The prototype galaxy is NGC 891, which has been extensively studied at all wavelengths \citep{Rand.1998, Hodges.and.Bregman.2013, Seon.etal.2014, Hughes.etal.2015}. Many studies have emphasized that the observed properties of the extraplanar DIG in edge-on galaxies cannot be purely due to  Lyman photons escaping from disc \hii regions. A variety of suggestions have been put forward: dissipation of turbulence \citep{Minter.and.Spangler.1997}, magnetic reconnection \citep{Raymond.1992}, shocks \citep{Collins.and.Rand.2001}, cosmic rays, photoelectric heating from interstellar dust grains \citep{Weingartner.and.Draine.2001}, and Lyman photons from old stars \citep{FloresFajardo.etal.2011a}.

CALIFA data can bring new insight into this problem. Fig.\ \ref{fig:ExampleMapsEdgeOn} shows SDSS images and our CALIFA \Ha maps for five edge-on galaxies in our sample.\footnote{While CALIFA 0077, 0936, and 0811 are very nearly edge-on, CALIFA 0822 and 0791 have inclinations of $\sim 59^\circ$ and $65^\circ$ respectively. The latter two galaxies are nevertheless still useful in this analysis as long as one considers locations far from the disc plane.} The layout is as in Fig.\ \ref{fig:ExampleMaps}. The top four galaxies show a very similar structure in their $W_{\Ha}$ maps, with SFc emission concentrated in the disc, where it is surrounded by mDIG. Above and below the plane, however, essentially all emission is hDIG-like. This provides strong support for the scenario of \citet{FloresFajardo.etal.2011a} where the ionization becomes dominated by HOLMES at large galactic latitudes. Maps of standard diagnostic line ratios reinforce this conclusion, as illustrated in the MaNGA-based studies by \citet{Belfiore.etal.2016} and \citet{Zhang.etal.2017a}.

For the first time, we can relate this extraplanar DIG emission to the underlying stellar population. We do this by using the {\sc starlight} results to compute the ratio $\xi$ (see Section \ref{sec:hDIG}). For hDIG regions in the top four edge-on galaxies in Fig.\ \ref{fig:ExampleMapsEdgeOn}, we find a median $\xi$ value of 1.5, and an interquartile range from 1.1 to 1.9. Given the factor of $\sim 2$--3 uncertainty in this estimate (CF11), the main conclusion here is that $\xi$ is of the order of 1, and thus that the old stellar populations in these regions produce enough $h\nu > 13.6$ eV photons to explain the observed extraplanar \Ha emission.

Face-on views of these same galaxies would project the extraplanar hDIG on top of a predominantly SFc + mDIG disc. For a constant \Ha emissivity, the ratio of face-on to edge-on $\Sigma_{\Ha}$ equals the ratio $h / r$ between height and radius of the extraplanar hDIG layer. In the edge-on examples in Fig.\ \ref{fig:ExampleMapsEdgeOn}, the extraplanar hDIG has $\Sigma_{\Ha}$ of the order of a few times $10^4 L_\odot\,$kpc$^{-2}$. For $h \sim r$, this is also the expected face-on surface brightness of this component. This is much smaller than the $\Sigma_{\Ha}$ of SFc in the $\sim$ face-on systems shown in Fig.\ \ref{fig:ExampleMaps}, so that the projected extraplanar emission has a negligible effect. Some mDIG regions, however, have $\Sigma_{\Ha}$ values not much larger than $10^4 L_\odot\,$kpc$^{-2}$, and thus could carry a non-negligible contribution from extra-planar hDIG.

The galaxy in the bottom row (CALIFA 0811, UGC 10043) differs from the others in Fig.\ \ref{fig:ExampleMapsEdgeOn} in having much more SFc along its disc, as well as SFc-like extra-planar emission over the disc, and a bipolar cone of intermediate $W_{\Ha}$ values centred in the nucleus. This galaxy has been recently studied by \citet{LopezCoba.etal.2017}, who find the emission-line ratios and kinematics along the biconical structure to be consistent with a galactic wind powered by a central SF event. This combination of shock ionization and the widespread SF along its disc explains why there are so little signs of hDIG emission in this galaxy, although it is curious that $W_{\Ha}$ does drop close to hDIG values in the inner parts of the bicone.

\subsection{Comparison with $\Sigma_{\Ha}$-based SF/DIG separation schemes}
\label{sec:SBHa}

%---------------------------- Figure ----------------------------
\begin{figure}
 \includegraphics{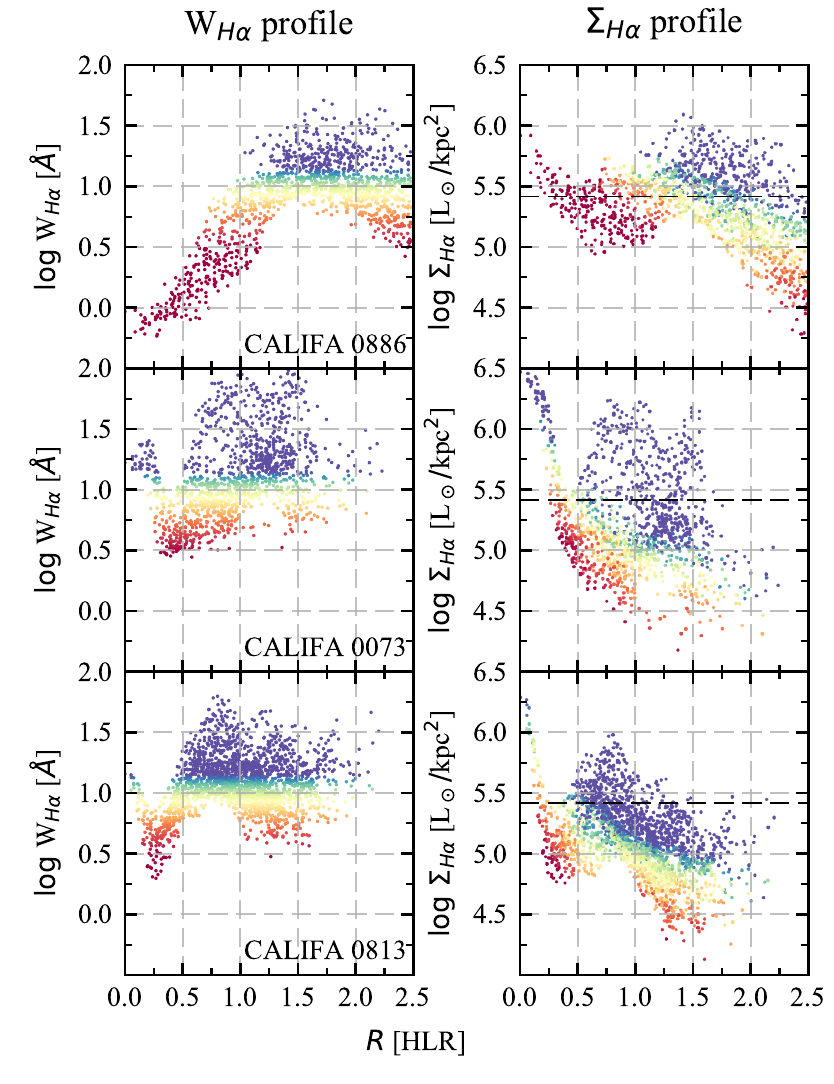}
 \caption{
  $W_{\Ha}$ and $\Sigma_{\Ha}$ profiles for three of the example galaxies in Fig.\ \ref{fig:ExampleMaps}. Points are coloured according to $W_{\Ha}$. The dashed lines in the right-hand panels mark $\Sigma_{\Ha} = 10^{39}$ erg$\,$s$^{-1}\,$kpc$^{-2}$.
 }
 \label{fig:WHa_and_SHa_profiles}
\end{figure}
%---------------------------- Figure ----------------------------

Despite its conceptual advantages insofar as distinguishing different nebular regimes is concerned, $W_{\Ha}$ contains $\Sigma_{\Ha}$ in its numerator, so  one may think that criteria based on these two variables may end up producing similar results. The maps in Fig.\ \ref{fig:ExampleMaps}  show that structures like SF arms are indeed similarly traced by $W_{\Ha}$ and $\Sigma_{\Ha}$, but others are not. Most notably, $\Sigma_{\Ha}$ \textit{always} peaks in the central regions of galaxies while, for early-type galaxies, $W_{\Ha}$ shows clear dips.

Fig.\ \ref{fig:WHa_and_SHa_profiles} examines this issue by means of radial profiles for CALIFA 0886, 0073, and 0813, three of the examples in Fig.\ \ref{fig:ExampleMaps}. (For more examples of $W_{\Ha}$ profiles see \citealt{Papaderos.etal.2013, Belfiore.etal.2016, Belfiore.etal.2017, Gomes.etal.2016b, GonzalezDelgado.etal.2016a}.) The left- and right-hand panels plot $W_{\Ha}$ and $\Sigma_{\Ha}$ against $R$, respectively. Both are coloured by $W_{\Ha}$ using the same colour scheme of previous plots.

CALIFA 0886 is a good example of the many galaxies that show low-$W_{\Ha}$, hDIG-dominated emission in their central regions, yet a peak in $\Sigma_{\Ha}$. The reason why these retired bulges appear brighter than the surrounding disc is the much higher concentration of HOLMES in the bulge. This geometrically enhanced \Ha emission can be erroneously attributed to SF with an SF/DIG criterion based on $\Sigma_{\Ha}$. As seen in the top-right panel, the inner regions of this galaxy cross the $\Sigma^{\rm SF,min}_{\Ha} = 10^{39}$ erg$\,$s$^{-1}\,$kpc$^{-2} =  2.6 \times 10^{5} L_\odot\,$kpc$^{-2}$ threshold claimed to `select reliable \hii region dominated spaxels' according to \citet{Zhang.etal.2017a}. Yet, their $W_{\Ha}$ of $\sim 1$ \AA\ are firmly in the hDIG regime. The $W_{\Ha}$ criterion thus correctly identifies the bulge of this and other galaxies as retired, while a  $\Sigma_{\Ha}$ criterion would interpret them as bright, SF regions.

Throughout most of the disc of CALIFA 0886, the \citet{Zhang.etal.2017a} $\Sigma_{\Ha}$-based SF/DIG criterion agrees with the nebular regime identified via $W_{\Ha}$. This agreement is only partial in CALIFA 0073 (central panels in Fig.\ \ref{fig:WHa_and_SHa_profiles}), where we find more disc SF regions with $W_{\Ha}$ than with $\Sigma_{\Ha}$, and even more so in CALIFA 0813 (bottom panels), where most $W_{\Ha} > 14$ \AA\ zones  fall below the $\Sigma^{\rm SF,min}_{\Ha}$ threshold. These differences stem from the contrasting radial behaviours of $\Sigma_{\Ha}$ and $W_{\Ha}$. For a fixed $W_{\Ha}$, the highest $\Sigma_{\Ha}$'s tend to be located in central regions, while, in contrast, for a fixed $\Sigma_{\Ha}$, the largest values of $W_{\Ha}$ are generally found in the outskirts. Indeed, as seen in the examples in Fig.\ \ref{fig:WHa_and_SHa_profiles}, $\Sigma_{\Ha}$ tends to decrease outwards while $W_{\Ha}$ remains roughly constant, both with large dispersions at any given $R$ in the disc. About 37 per cent of our SFc spaxels have $\Sigma_{\Ha} < 10^{39}$ erg$\,$s$^{-1}\,$kpc$^{-2}$. On average, these faint SF regions are at $R = 1.3$ HLR from the centre.

In summary, compared to the hDIG/mDIG/SFc separation criteria proposed in this paper, a $\Sigma_{\Ha}$-based criterion tends to overestimate the population of SF regions at low $R$ and underestimate it at large $R$.  More worryingly, as already mentioned,   $\Sigma_{\Ha}$ by itself cannot identify the hDIG component, the main source of \Ha emission in old spheroids. In fact, we have seen that retired bulges may be mistaken for their very opposite when bright enough to exceed $\Sigma^{\rm SF,min}_{\Ha}$.

Previous CALIFA-based studies by \citet{Kehrig.etal.2012}, \citet{Singh.etal.2013}, and \citet{Gomes.etal.2016b} also find $\Sigma_{\Ha}$ values in excess of the $\Sigma^{\rm SF,min}_{\Ha}$ limit of \citet{Zhang.etal.2017a} in the inner regions of early-type galaxies, where the absence of young stars is beyond dispute (see also \citealt{Sarzi.etal.2010} for results based on the SAURON survey). These examples are the observational realization of the DIG + DIG = SF conceptual inconsistency pointed out in Section \ref{sec:WHaRationale}. The $W_{\Ha}$-based scheme presented in this paper solves this problem by extending to spatially resolved data the concept of retired galaxy proposed by \citet{Stasinska.etal.2008a} and CF11 in the context of integrated galaxy spectra.

\subsection{Can the DIG be detected with density-sensitive line ratios?}
\label{sec:nSii}

\begin{figure}
 \includegraphics{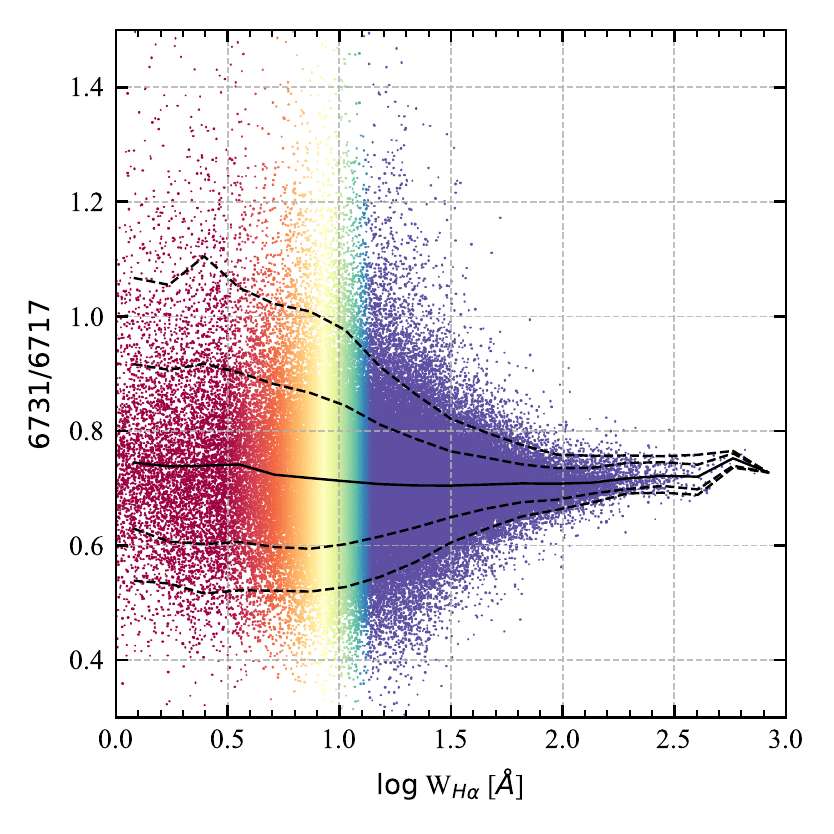}
 \caption{[S\,{\sc ii}]$\lambda\lambda$6731/6716 flux ratio for 111760 zones in our sample where this ratio has $SN > 3$. Points are coloured by $W_{\Ha}$ as in previous figures. The solid line traces the median curve, and dashed lines show the $1\sigma$ and $2\sigma$ equivalent percentiles.}
 \label{fig:S2_x_WHa}
\end{figure}

The electron densities in the Milky Way DIG obtained from dispersion measures and \hi column densities towards pulsars at known distances are typically less than $10^{-1}$ cm$^{-3}$ \citep{Berk.and.Fletcher.2008},  orders of magnitude smaller than those of \hii\ regions. One could then expect that the density-sensitive \Sii\ line ratios in our data would indicate smaller densities in the DIG than in  SFc, and show a tendency with $W_{\Ha}$. Fig.\ \ref{fig:S2_x_WHa} plots the \sii 6731/6716 flux ratio as a function of $W_{\Ha}$ (only zones where $SN > 3$ in the \sii ratio are plotted). The median and the $\pm 1\sigma$ and $2\sigma$ percentiles curves are overplotted. We find no trend of  \sii 6731/6716 with $W_{\Ha}$. The increase in scatter towards lower $W_{\Ha}$ is consistent with the corresponding decrease in the $SN$ of the lines. In essence, \sii 6731/6716, in regions where this ratio can be safely measured, is everywhere consistent with its low density limit of 0.7. Of course, this figure cannot say anything about regions where $SN < 3$.

The reason why the \sii line ratio is not substantially different in the DIG zones where it can be measured and in the SFc is twofold: first, at our resolution, SFc contain a significant amount of diffuse gas. Secondly, the \sii line ratio is not sensitive to densities below $\sim 50$\,cm$^{-3}$.

One could expect to obtain a different picture if using a line doublet sensitive to lower electron densities, like the far-infrared doublet [N\,{\sc ii}]$\lambda\lambda$205, 125$\mu$m. Recent observations have allowed one to map this ratio in the Milky Way and several galaxies \citep{Goldsmith.etal.2015, HerreraCamus.etal.2016}. The derived densities range between 1 and 300 cm$^{-3}$, with a median value of 30 cm$^{-3}$, and do not get anywhere close to the DIG densities obtained from pulsar measurements.

Classical density estimators are thus not able to detect the DIG, at least not at the resolution of CALIFA and similar surveys. However, it is not clear whether they would do a better job at higher spatial resolution, since, as noted by \citet{Rubin.1989} in another context, density inhomogeneities severely hamper any quantitative interpretation of such density-sensitive line ratios.

%---------------------------- Figure ----------------------------
%---------------------------- Figure ----------------------------

\subsection{$W_{\Ha}$ and the BPT diagram}
\label{sec:BPT}

%---------------------------- Figure ----------------------------
\begin{figure*}
 \includegraphics{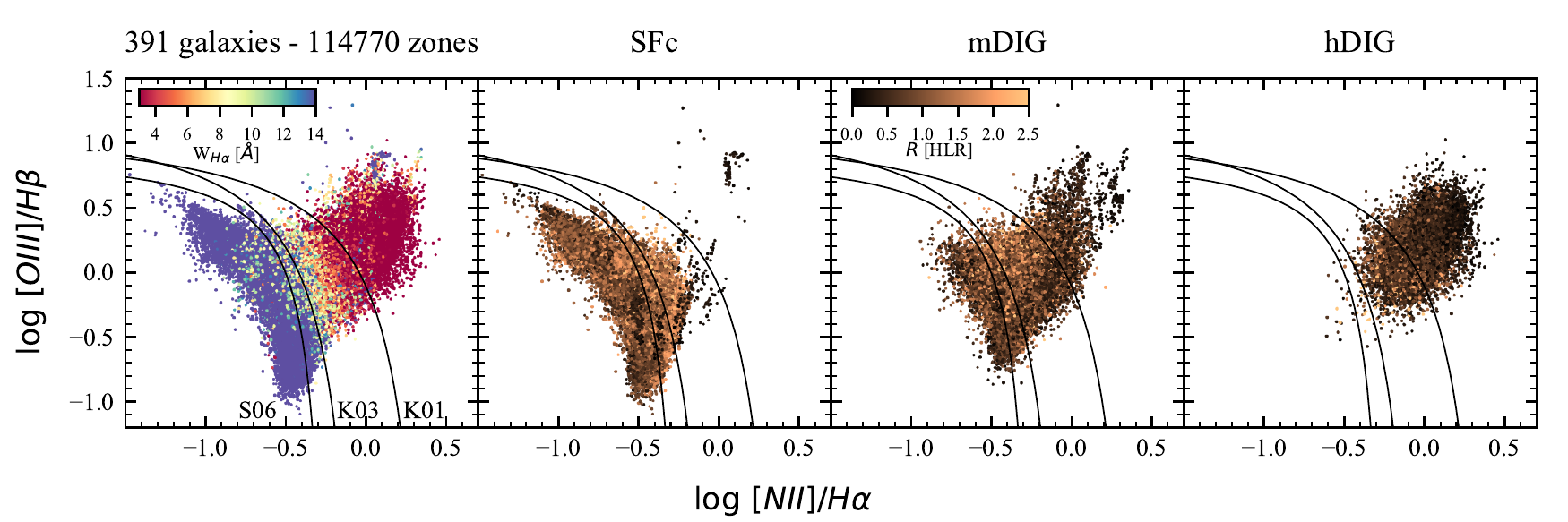}
 \caption{BPT diagrams for our sample. The left-hand panel shows the full sample, with colours coding for $W_{\Ha}$, as indicated. Other panels split the sample into SFc ($W_{\Ha} > 14$ \AA), mDIG ($W_{\Ha} = 3$--14 \AA), and hDIG ($W_{\Ha} < 3$ \AA) regions, colouring points according to their radial distance $R$ (in HLR units).  In all cases, only zones with $SN > 3$ in all lines are plotted. Dividing curves come from (from left to right) \citet[S06]{Stasinska.etal.2006a}, \citet[K03]{Kauffmann.etal.2003a}, and \citet[K01]{Kewley.etal.2001a}.
 }
 \label{fig:BPT}
\end{figure*}
%---------------------------- Figure ----------------------------

The different ionization and heating conditions in hDIG, mDIG, and SFc should lead to different collisional-to-recombination line flux ratios, and thus to different loci on diagnostic diagrams like \Oiii/\Hb versus\ \Nii/\Ha. This famous BPT diagram \citep*[after][]{Baldwin.Phillips.Terlevich.1981a} is widely used to separate SF galaxies from those where harder ionizing sources contribute significantly to the ionization of the gas. It is thus legitimate to ask whether our $W_{\Ha}$-based scheme is consistent with this independent way of characterizing the nebular regime.

Fig.\ \ref{fig:BPT} shows BPT diagrams obtained from all zones in our sample where $SN > 3$ in all four lines involved. The top-left panel shows results for the entire sample, colouring points according to $W_{\Ha}$ following the scheme used for the right-hand panels in Fig.\ \ref{fig:ExampleMaps}, which saturates at $> 14$ \AA\ (blue) and $< 3$ \AA\ (red). The curves correspond to the demarcation lines proposed by \citet[S06]{Stasinska.etal.2006a}, \citet[K03]{Kauffmann.etal.2003a}, and \citet[K01]{Kewley.etal.2001a} -- see CF11 for a discussion of the meaning of these curves.

The strong correspondence between $W_{\Ha}$ and the BPT coordinates is evident to the eye, as previously noted by \citet{Morisset.etal.2016} for CALIFA data and \citet{Belfiore.etal.2016} for MaNGA. The left wing is predominantly populated by SFc, while hDIG regions populate the tip of the right wing, with mDIG regions in between. This correspondence is further confirmed in the right-hand panels in Fig.\ \ref{fig:BPT}, which separate SFc, mDIG, and hDIG zones according to our criterion, with points coloured by their radial distance to the nucleus. Deviant points are concentrated in the inner regions of galaxies (red--orange points), where some high-$W_{\Ha}$ SFc or mDIG zones intrude into a region in the BPT diagram otherwise dominated by hDIG emission. These outliers come from AGN in our sample, as discussed in Section \ref{sec:Caveats}.

Fig.\ \ref{fig:BPT} thus makes a visually compelling case for our $W_{\Ha}$-based SFc/mDIG/hDIG separation scheme. Translating this impression to numbers, and restricting the analysis to $R > 1$ HLR, 58 (92)  per cent of our $W_{\Ha} > 14$ \AA\ zones are in the SF region of the BPT diagram according to the S06 (K03)
criterion. The relatively large number of zones trespassing the S06 line is not surprising if one recalls that this curve was derived on the basis of photoionization models designed to establish the boundaries of \textit{pure} SF regions in the BPT plane. As reiterated several times in this paper, at the resolution of CALIFA, our SFc are nowhere near pure \Hii\ regions, but include plenty of diffuse emission, which enhances both line ratios in the BPT diagram.

We thus conclude that our SFc/mDIG/hDIG separation scheme leads to diagnostic line ratios that are qualitatively consistent with what one expects for these nebular regimes. In conjunction with the conceptual and empirical arguments presented in Section \ref{sec:DIGxHII}, this adds further strength to our methodology.

\subsection{The mDIG as an SF+hDIG mixture}
\label{sec:TheNatureOfTheMIG}

%---------------------------- Figure ----------------------------
\begin{figure}
 \includegraphics{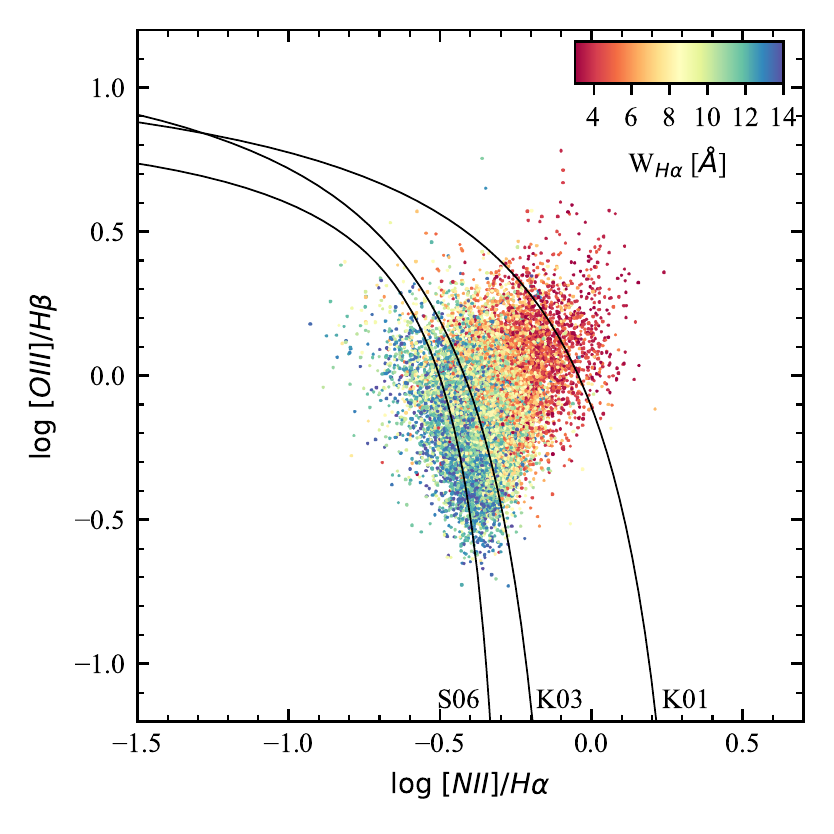}
 \caption{BPT diagram for mDIG regions only (i.e., those with $W_{\Ha}$ in the 3--14 \AA\ range), coloured according to $W_{\Ha}$, and excluding zones inwards of $R = 1$ HLR.}
 \label{fig:BPT_MIG}
\end{figure}
%---------------------------- Figure ----------------------------

Fig.\ \ref{fig:BPT} shows that mDIG zones straddle the region between the classical SF wing and the loci of hDIG zones in the BPT diagram, a behaviour that lends itself to an interpretation in terms of a mixture of SFc and hDIG emission. This behaviour is better appreciated in Fig.\ \ref{fig:BPT_MIG}, which shows the mDIG points coloured according to $W_{\Ha}$. To mitigate potential contamination by AGN-affected zones, only points at $R > 1$ HLR ($= 5.3$ kpc on average) are shown.

The same steady progression in $W_{\Ha}$  along the right wing seen in the top-left panel of Fig.\ \ref{fig:BPT} is also seen in Fig.\ \ref{fig:BPT_MIG}, suggesting that the mDIG component may be described as a mixture of SFc and hDIG emission. That the mDIG is interpretable as a mixture of processes is of course just what one would expect. The large population of mDIG zones approaching the $W_{\Ha} = 14$ \AA\ threshold  and whose BPT coordinates overlap with those of SFc probably correspond to the leakage scenario. Above the SF wing in the BPT diagram, the main ionization process is likely still due to young, massive stars, but the influence of the heating due to HOLMES increases gradually as $W_{\Ha}$ decreases. When $W_{\Ha}$ approaches the $< 3$ \AA\ hDIG regime,  the ionizing field of HOLMES starts dominating the photoionization budget too.

\subsection{AGN and other caveats}
\label{sec:Caveats}

Our whole hDIG/mDIG/SFc classification scheme ignores other mechanisms of line production, most notably AGN. AGN are found in the central parts of some galaxies and recognized as such through the BPT diagram. For example, the cluster of blue ($W_{\Ha} > 14$ \AA) points at the tip of the right wing of the BPT in Fig.\ \ref{fig:BPT}, at coordinates $\sim (0.2,0.8)$ come from the central regions of CALIFA 0897 (UGC12348),  a known type 2 Seyfert \citep{Cusumano.etal.2010, Asmus.etal.2014}. Other $W_{\Ha} > 3$ \AA\ outliers in BPT loci otherwise dominated by hDIG zones also tend to be located at small $R$ (reddish points in the central panels of Fig.\ \ref{fig:BPT}).

AGN may also power line emission well outside the nucleus \citep[up to distances as large as 20 kpc;][]{Veilleux.etal.2003}. These are the so-called extended emission-line regions (EELRs) or ionization cones. They can be due either to photoionization by X-ray photons leaving the AGN with a small opening angle or to an interaction between radio jets and the galaxy ISM producing strong shocks \citep{Wilson.1996}. However, in the framework of the present study, which is to evaluate the importance of the DIG in galaxies and pinpoint its different regimes, EELRs in Seyfert galaxies are a secondary issue, as they affect only specific zones of galaxies with a well-defined AGN -- and perhaps not all of them. Understanding the EELRs is a topic in itself, which indeed can be tackled with sensitive 3D spectroscopy, and some recent studies already started doing so \citep[e.g.,][]{Dopita.etal.2014}, but it is outside the scope of the present paper.

Another line-producing process neglected in this study is shocks. In the case of the galactic wind in CALIFA 0811, shown in Fig.\ \ref{fig:ExampleMapsEdgeOn}, we find $W_{\Ha} = 3$--12 \AA\ in the shocked region, i.e., mDIG-like values. Again, $W_{\Ha}$ by itself cannot identify the shock origin of the nebular emission, though it at least tells that photoionization by HOLMES is not an energetically feasible explanation. Only a detailed study of the geometry, line ratios, and kinematics can fully reveal the processes governing line emission in objects like this \citep{Kreckel.etal.2014, Beirao.etal.2015, LopezCoba.etal.2017}.

Because of their relative rarity and spatial constraints, these processes do not affect much the hDIG/mDIG/SFc statistics found in this study. They should nevertheless be taken into consideration in studies of individual sources.

Finally, a word of caution about the so-called composite region in the BPT diagram, commonly defined as the zone below the K01 line and above the K03 or S06 line. Though this region is usually thought to correspond to SF + AGN mixtures, AGN and hDIG have indistinguishable line ratios. It is therefore not a priori clear what this composite region is composed of!

The way to break this degeneracy is through $W_{\Ha}$. Given that (unlike AGN or SF) old stars are everywhere in galaxies, one should understand the hDIG regime as a floor level of ionization, one that is only energetically relevant  when no other source is. Whenever the stellar continuum around \Ha is dominated by old stars (which is  the case even in SF regions at  kpc resolution), the direct scaling between the optical continuum and the ionizing fluxes of the corresponding HOLMES population leads to $W_{\Ha} \sim 1$--2 \AA\ according to current evolutionary population synthesis models (see CF11 and references therein). Thus, composite spectra where $W_{\Ha}$ approaches this limit likely represent an SFc + hDIG mixture. Conversely, as long as $W_{\Ha}$ is above the hDIG range, an SF + AGN mixture is more plausible.

The \citet{Davies.etal.2014} study illustrates this point. Using the CALIFA data cubes of two type 2 Seyferts (NGC 2410 and NGC 6394) and two other more ambiguous (Seyfert--LINER) cases (IC 0540 and NGC 6762), they identify approximately one-dimensional distributions in the BPT and other diagnostic diagrams, suggestive of SF + AGN mixing sequences. They however verified that in NGC 6762 the majority ($> 90 per cent$) of the spaxels have $W_{\Ha} < 3$ \AA, so that the contribution of HOLMES cannot be ignored. NGC 2410 and NGC 6394, on the other hand, have central $W_{\Ha}$ values well above the hDIG range, which makes the SF + AGN interpretation plausible. IC0540 has $W_{\Ha}$ values in the mDIG-to-hDIG range, so that the interpretation is less clear, although they favour an SF + AGN scenario. The bottom line here is $W_{\Ha}$ should be taken into account when studying composite systems in order to avoid confusing hDIG and AGN effects.

\section{Summary}
\label{sec:Conclusions}

We have used data cubes from the CALIFA survey to investigate the origin of the line emission in over 300k zones of 391 galaxies across the Hubble sequence. Studies based on integrated galaxy data like the SDSS give great emphasis on determining whether or not a galaxy hosts an AGN. In a sample of spaxel spectra like ours, however, a more relevant question is whether the ionization is locally dominated by photons arising from massive stars (either in bona fide \hii\ regions or in diffuse regions surrounding them) or by photons arising from old stellar populations. This has been the focus of our study.

We have shown that the commonly adopted method to separate DIG from SF regions on the basis of the surface brightness of \Ha ($\Sigma_{\Ha}$) is conceptually flawed, and that the \Ha equivalent width ($W_{\Ha}$) is more suitable to distinguish the qualitative differences between these regimes. Moreover, and perhaps more importantly, $W_{\Ha}$ further allows us to confidently identify the cases where the line emission is predominantly powered by HOLMES, an omnipresent population that provides a floor level of ionizing radiation in galaxies.

The observed distribution of $W_{\Ha}$ within and among galaxies was used to propose a classification scheme. Zones where $W_{\Ha} < 3$ \AA\ are attributed to a HOLMES-ionized gas (hDIG) component, responsible for the $W_{\Ha} \sim 1$ \AA\ peak in the strongly bimodal distribution of $W_{\Ha}$. This observational definition of hDIG is identical to the one for retired galaxies. In the absence of any (astro)physically motivated argument, we have defined the concept of SFc as regions where $W_{\Ha} > 14$ \AA, the mode of the high-$W_{\Ha}$ population in our sample. Regions where $W_{\Ha}$ falls in the 3--14 \AA\ intermediate range are tagged as having a mixed nature (mDIG).

Some of the main results obtained with this empirically and theoretically motivated scheme are as follows.

\begin{enumerate}

\item In agreement with their predominantly old stellar populations, the hDIG is the main nebular regime in early-type galaxies (E and S0) and in bulges.

\item The extraplanar emission in edge-on spirals is also of hDIG type, vindicating the scenario put forward and elaborated by \citet{FloresFajardo.etal.2011a}. In face-on systems, this extraplanar hDIG makes a negligible contribution when projected over SFc, but a potentially relevant one when projected on to regions classified as mDIG.

\item A $\Sigma_{\Ha}$-based SF/DIG separation scheme tends to classify hDIG-dominated retired bulges as SF, an inconsistency that is ultimately due to the extensive nature of $\Sigma_{\Ha}$. In addition, faint SF regions tend to be misclassified as DIG with a $\Sigma_{\Ha}$ criterion.

\item The hDIG, mDIG, and SFc contributions to the \Ha luminosity vary in a systematic way along the Hubble sequence, ranging from $(100, 0, 0)$ per cent in ellipticals and S0's to $(9, 60, 31)$ per cent in Sa--Sb's and $(0, 13, 87)$ per cent in later types.

\item SFc and hDIG regions occupy characteristic loci on the BPT diagram. mDIG regions form a continuous sequence between SFc and hDIG line ratios, indicative of an mDIG = SFc + hDIG mixture.

\end{enumerate}

%***************************************************************%
%                                                               %
%                       Acknowledgments                         %
%                                                               %
%***************************************************************%

\section*{ACKNOWLEDGEMENTS}
\noindent EADL, RCF, GS, and NVA acknowledge the support from the CAPES CsF--PVE project 88881.068116/2014-01. RGD acknowledges the support of CNPq (Brazil) through Programa Ci\^encia sem Fronteiras (401452/2012-3). CALIFA is the first legacy survey carried out at Calar Alto. The CALIFA collaboration would like to thank the IAA-CSIC and MPIA-MPG as major partners of the observatory, and CAHA itself, for the unique access to telescope time and support in manpower and infrastructures. We also thank the CAHA staff for the dedication to this project. Support from the Spanish Ministerio de Econom\'\i a y Competitividad, through projects AYA2016-77846-P, AYA2014-57490-P, AYA2010-15081, and Junta de Andaluc\'\i a FQ1580.

\bibliographystyle{mnras}
\bibliography{bibliografia}

\bsp
\label{lastpage}
\end{document}